\documentclass[aps,pre,twocolumn,floatfix,nofootinbib]{revtex4}

\usepackage{amssymb,amsfonts,amsmath}
\usepackage{graphicx}
\usepackage{dcolumn}
\usepackage{bm}
\usepackage{mathrsfs}
\usepackage{mhchem}
\usepackage{subfigure}
\usepackage{color}
\usepackage{soul}

\usepackage{upgreek}

\usepackage{booktabs}
\usepackage{epstopdf}
\usepackage{multirow}

\usepackage{chemarrow}
\usepackage{extarrows}
\usepackage{mathtools}

\usepackage{epsf}
\usepackage{epstopdf}
\newcommand{\be}{\begin{equation}}
\newcommand{\ee}{\end{equation}}
\newcommand{\bea}{\begin{eqnarray}}
\newcommand{\eea}{\end{eqnarray}}

\usepackage[T3,T1]{fontenc}
\DeclareSymbolFont{tipa}{T3}{cmr}{m}{n}
\DeclareMathAccent{\invbreve}{\mathalpha}{tipa}{16}

\usepackage{enumitem}
\setlist[itemize]{topsep=0.05cm}

\usepackage{lipsum}

\begin{document}

\setlength{\abovedisplayskip}{5pt}
\setlength{\belowdisplayskip}{5pt}
\setlength{\abovedisplayshortskip}{5pt}
\setlength{\belowdisplayshortskip}{5pt}

\title{From the thermodynamics of irreversible processes \\ to dissipative structures and active matter}

\author{Pierre Gaspard}
\thanks{ORCID: {\tt 0000-0003-3804-2110}}
\email{Gaspard.Pierre@ulb.be}
\affiliation{ Center for Nonlinear Phenomena and Complex Systems, Universit{\'e} Libre de Bruxelles (U.L.B.), Code Postal 231, Campus Plaine, B-1050 Brussels, Belgium}


\begin{abstract}
{\bf Abstract.} A historical perspective is presented on thermodynamics from the pioneering contributions by Carnot and Clausius to recent advances on active matter.  Non-equilibrium thermodynamics develops from the identification of the irreversible processes contributing to entropy production in various types of materials and systems.  These processes include friction, viscosity, heat and electric conductions, diffusion, reactions, and more.  In 1954, Glansdorff and Prigogine formulated a general evolution criterion, which led to the theory of dissipative structures like chemical clocks, reaction-diffusion patterns, and convection patterns.  Non-equilibrium statistical mechanics provides the microscopic foundations for the thermodynamics of irreversible processes.

\end{abstract}


\maketitle

\vspace{-0.8cm}

\section{Historical introduction}
\label{sec:intro}

\vspace{-0.2cm}

In 1824, Sadi Carnot published his memoir entitled ``R\'eflexions sur la puissance motrice du feu'', a study of heat engines, which inspired the formulation of the second law of thermodynamics \cite{C1824}.  The historical context was the development of steam engines (external combustion engines) for industry, mining, and transportation by pioneers like Papin, Savery, Newcomen, Cugnot, Watt, Trevithick, and others; the major issue being the conversion of heat into mechanical work.  In his memoir, Sadi Carnot also mentioned that heat is the cause of movements in the atmosphere, oceans, and volcanoes, already opening the scope of thermodynamics to the study of natural phenomena.  In his reasonings, he used the laws of ideal gases by Mariotte (1676, $p\propto V^{-1}$) and Gay-Lussac (1802, $p\propto T$) to obtain the following theorem: ``In all cases where a quantity of heat is transformed into work, and the body effecting this transformation finally returns to its initial state, another quantity of heat must necessarily pass from a warmer to a colder body, and the magnitude of the latter quantity of heat in relation to the former depends only on the temperatures of the two bodies between which it passes, and not on the nature of the body effecting the transformation'', as stated by Clausius in 1854.  This result led Clausius to shortly formulate the second law of thermodynamics as: ``Heat cannot pass by itself from a cold body to a warmer body'' \cite{C1854,C1867}.

Today, we know many more forms of energy than only heat and work:

\begin{itemize}
\setlength\itemsep{-0.2em}

\item{Chemical energy:} e.g. in internal combustion engines for boats, trains, cars, aircrafts \& rockets; or
other processes in chemistry \& biochemistry;

\item{Electric energy:} e.g. in electrochemical processes, electric batteries, electric engines, thermoelectricity, plasmas \& microelectronics;

\item{Electromagnetic energy:} e.g. in electromagnetic waves, light, lasers \& harvested by photovoltaic cells;

\item{Nuclear energy:} e.g. in nuclear fusion (the energy source of stars) \& nuclear fission;

\item{Gravitational energy:} e.g. in supernovae \& black holes.

\end{itemize}

Again, the major issue is the conversion of energy from one form into another.  Nowadays, this issue also concerns the biosciences (biochemistry, biophysics, and molecular biology), surface science (including heterogeneous catalysis), the nanosciences, and information science.  In all these fields, thermodynamics is needed to understand the mechanisms of energy conversion and their efficiencies, given the limits arising from energy conservation and the non-negativity of entropy production.

In the earliest developments of thermodynamics by Carnot, Clausius, and others during the XIXth century, the first and second laws of thermodynamics were enunciated in a global formulation, defining energy and entropy for an entire system, possibly in contact with other ones (see Sec.~\ref{sec:thermo-laws}).  Irreversible processes contributing to entropy production were discovered in different kinds of materials and systems (see Sec.~\ref{sec:IrrevProc}).  Many irreversible processes such as viscosity, heat conduction, or diffusion are intrinsic local properties of fluids and other continuous media, which gave rise to the local formulation of thermodynamics by Natanson \cite{N1896}, Duhem \cite{D1911}, Jaumann \cite{J1911,J1918}, Lohr \cite{L1916,L1924}, and others during the period extending from the 1890s to the 1920s.  In the 1930s, the contribution of chemical reactions to entropy production was established by De~Donder \cite{D1936}.  Furthermore, Onsager proved his famous reciprocal relations on the basis of microreversibility \cite{O31a,O31b}.  Afterwards, a great theoretical synthesis was carried out in the early 1940s with the works of Meixner \cite{M1939-1943}, Eckart \cite{E1940a,E1940b,E1940c}, and Bridgman \cite{B1940}; and after WWII with those of Prigogine \cite{P47,P55}, de~Groot \cite{G51}, de~Groot \& Mazur \cite{GM62}, Callen \cite{C65}, and others on the non-equilibrium thermodynamics of fluids, solutions and electrolytes, which are composed of inert or reactive species (see Sec.~\ref{sec:noneq-thermo}).  Later, the theory was extended to the phases of matter with broken continuous symmetries like crystals, liquid crystals, and superfluids \cite{MPP72}.  

In the 1950s and 1960s, the thermodynamics of open systems exchanging energy, matter, and entropy with their environment knew significant advances, proving that the entropy produced by irreversible processes in the interior of the system can be evacuated to the exterior, possibly reducing the entropy inside the system \cite{P55}.  In this way, thermodynamics was shown to be compatible with the formation of patterns and oscillations -- called dissipative structures -- under far-from-equilibrium conditions \cite{PN67,PL68,NP77}.  These conditions require that the non-equilibrium constraints should exceed some threshold in order for dissipative structures to emerge, which is in relation to the general evolution criterion of Glansdorff \& Prigogine \cite{GP54,GP71}, as briefly explained in Sec.~\ref{sec:DS-GEC}.  Since then, many spatiotemporal dissipative structures have been discovered and characterized in physics, chemistry, the biosciences, the geosciences, and astronomy.

Recently, research has focused on active matter, which is made of self-propelled particles or other agents locally converting energy into mechanical motion \cite{MJRLPRA13,K13,BDLRVV16}.  Active matter therefore operates out of equilibrium and its non-equilibrium thermodynamics has become a major topic of interest \cite{JGS18,GK19,GK20,MFTC21,ATR25,BRS25,G25}, as shortly presented in Sec.~\ref{sec:AM}.

In parallel with these developments, much effort has been devoted to establishing the microscopic foundations of thermodynamics, which is summarized in Sec.~\ref{sec:MicroFoundations}.  Conclusion and perspectives are given in Sec.~\ref{sec:conclusion}.

\vspace{-0.3cm}

\section{The laws of thermodynamics}
\label{sec:thermo-laws}

\vspace{-0.3cm}

\subsection{Global formulation}

\vspace{-0.3cm}

\begin{figure}[h]
\centerline{\scalebox{0.4}{\includegraphics{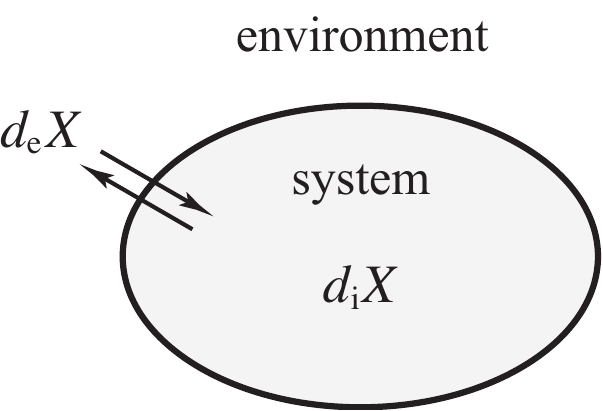}}}
\vspace{-0.2cm}
\caption{Schematic representation of an open system in contact with its environment.  $d_{\rm e}X$ is the contribution of exchanges and $d_{\rm i}X$ is the internal contribution to the differential $dX=d_{\rm e}X+d_{\rm i}X$ of some state variable $X$.}
\label{fig1}
\end{figure}

In the earliest formulations of thermodynamics, the state variables are defined globally for an entire system, which is possibly open exchanging energy and matter with its environment (see Fig.~\ref{fig1}).  Every state variable $X$ can change in time due to exchanges $d_{\rm e}X$ with the exterior of the system (i.e., its environment) and due to processes taking place in the interior of the system $d_{\rm i}X$.  Therefore, the differential of the state variable can be written as $dX=d_{\rm e}X+d_{\rm i}X$ \cite{P47,P55}.  If the state variable obeys a conservation law, the internal contribution is equal to zero, $d_{\rm i}X=0$, and $dX=d_{\rm e}X$. 

\vspace{-0.3cm}

The first law is the conservation of energy, which can thus be expressed as
\be\label{1st_law}
\boxed{dE = d_{\rm e} E + d_{\rm i} E
\qquad\mbox{with}\qquad d_{\rm i} E=0} \, .
\ee
Similar expressions hold for the conservation of mass and linear momentum.  All these conservation laws are justified on the basis of classical or quantum mechanics, which rules the motion of the microscopic particles composing matter.

The second law concerns the entropy, so-called by Clausius in reference to ``internal transformation'' ($\upepsilon\upnu\uptau\uprho$o$\uppi\upeta$ in Greek) and denoted $S$ \cite{C1865,C1867}.  In general, the entropy is not conserved, its internal production being always non-negative:
\be\label{2nd_law}
\boxed{dS = d_{\rm e} S + d_{\rm i} S
\qquad\mbox{with}\qquad d_{\rm i} S \ge 0} \, .
\ee
We have that $d_{\rm i} S=0$ for reversible transformations, e.g., at equilibrium; and $d_{\rm i} S>0$ for irreversible transformations, i.e., out of equilibrium.

The integration of Eq.~(\ref{2nd_law}) over a noncyclic process $S_0\to S$ gives $\int dS=S-S_0$.  On the basis of Carnot's study, Clausius showed that the integral of entropy exchange can be expressed as $\int d_{\rm e}S=\int\frac{dQ}{T}$ in terms of the heat $dQ$ exchanged with the environment and the temperature $T$ of the system.  Moreover, Clausius introduced the concept of ``uncompensated transformation'' $N\equiv \int d_{\rm i}S$, postulating its non-negativity $N\ge 0$, which is the statement of the second law, and mentioning friction as contributing to $N>0$.  Accordingly, Clausius obtained the formula
\be
N= S - S_0 - \int \frac{dQ}{T} \ge 0 \, ,
\ee
which is equivalent to $\int d_{\rm i}S=\int dS-\int d_{\rm e}S \ge 0$, given by integrating Eq.~(\ref{2nd_law}) \cite{C1865,C1867}.

Later, De Donder used the concept of ``uncompensated heat'', defined as $dQ^{\rm n} \equiv T d_{\rm i}S \ge 0$ \cite{D1936}.

\vspace{-0.4cm}

\subsection{Local formulation}

\vspace{-0.3cm}

In continuous media, the quantities of interest are fields locally defined in every point of the system such as the densities $x({\bf r},t)$.  In this regard, the corresponding global state variable is obtained by integrating the density over the volume of the system according to $X(t)\equiv \int_V x({\bf r},t) \, d{\bf r}$.
In general, some density obeys a local balance equation of the following form:
\be\label{bal_eq_x}
\partial_t \, x + {\rm div} \, \boldsymbol{\jmath}_x =\sigma_x \, , 
\ee
where $\boldsymbol{\jmath}_x$ is the current density or flux and $\sigma_x$ the production rate density that are associated with the density~$x$.  Integrating the balance equation~(\ref{bal_eq_x}) over the volume $V$ of the system and using the divergence theorem, the balance equation of the global formulation is recovered as
\bea
&&\frac{dX}{dt} = \frac{d_{\rm e} X}{dt} + \frac{d_{\rm i} X}{dt} 
\qquad\mbox{with}\qquad
\frac{dX}{dt} = \int_V \partial_t \, x \, d{\bf r} \, , \quad\nonumber\\
&&\frac{d_{\rm e}X}{dt} = -\oint_{\partial V} \boldsymbol{\jmath}_x \cdot d\boldsymbol{\Sigma} \, , \quad\mbox{and}\quad
\frac{d_{\rm i}X}{dt} = \int_V \sigma_x \, d{\bf r} \, ,
\eea
$d\boldsymbol{\Sigma}$ denoting an element of surface area of the boundary between the system and its environment.  In this regard, boundary conditions should be defined, which possibly take into account the non-equilibrium constraints imposed to the system by its environment.

In the local formulation, the first law is the local conservation of energy expressed as
\be\label{local_1st_law}
\boxed{\partial_t \, \epsilon + {\rm div} \, \boldsymbol{\jmath}_\epsilon = \sigma_\epsilon \qquad\mbox{with}\qquad \sigma_\epsilon = 0}\, ,
\ee
where $\epsilon$ is the energy density such that the total energy of the system is given by $E=\int_V \epsilon \, d{\bf r}$.  Since $\sigma_\epsilon=0$, integrating Eq.~(\ref{local_1st_law}) leads to $\frac{d_{\rm i}E}{dt}=0$ in consistency with the global formulation~(\ref{1st_law}) of the first law.  Similar local conservation equations hold for other conserved quantities such as mass and linear momentum.

Next, the second law is expressed by the following local balance equation for entropy:
\be\label{local_2nd_law}
\boxed{\partial_t \, s + {\rm div} \, \boldsymbol{\jmath}_s =\sigma_s
\qquad\mbox{with}\qquad \sigma_s \ge 0}\, ,
\ee
where $s$ is the entropy density such that $S=\int_V s \, d{\bf r}$ and implying that $\frac{d_{\rm i} S}{dt} \ge 0$ in agreement with the global formulation~(\ref{2nd_law}) of the second law.

In this way, the first and second laws of thermodynamics can be formulated in continuous media like fluids or solutions.

\vspace{-0.4cm}

\section{Irreversible processes}
\label{sec:IrrevProc}

\vspace{-0.3cm}

\subsection{Discoveries of irreversible processes}

\vspace{-0.3cm}

Since the XIXth century, many irreversible processes contributing to entropy production were discovered either in bulk phases like fluids and solids, or at interfaces between two bulk phases:

\begin{itemize}
\setlength\itemsep{-0.2em}

\item Shear \& dilatational viscous pressures (Navier 1822, Lamb 1879): \\ $\boldsymbol{\Pi} = - \eta \left(\boldsymbol{\nabla}{\bf v}+\boldsymbol{\nabla}{\bf v}^{\rm T}-\frac{2}{3} \boldsymbol{\nabla}\cdot{\bf v}\, \boldsymbol{\mathsf 1}\right)- \zeta\, \boldsymbol{\nabla}\cdot{\bf v}\, \boldsymbol{\mathsf 1}$;

\item Heat conduction: Fourier's law (1822): \\ $\boldsymbol{\mathcal J}_q = - \kappa\boldsymbol{\nabla} T$;

\item Diffusion: Fick's law (1855): $\boldsymbol{\mathcal J}_k = - {\mathcal D}_k \boldsymbol{\nabla} n_k$;

\item Thermodiffusion (Ludwig 1856, Dufour 1873, Soret 1879);

\item Electric conduction: Ohm's law (1827): $\boldsymbol{\mathcal J}_{\rm e} = \boldsymbol{\sigma}\cdot\boldsymbol{\mathcal E}$; \\ $V=RI$;

\item Thermoelectricity (Seebeck 1821, Peltier 1834, Thomson 1851);

\item Chemical reactions (De Donder 1932);

\item Interfacial solid-solid friction (mentioned by Clausius 1865);

\item Interfacial fluid-solid sliding friction (Navier 1827, Maxwell 1879);

\item Interfacial heat resistance (Kapitza 1941);

\item Interfacial diffusion and reactions in heterogeneous catalysis;

\end{itemize}
where $\boldsymbol{\Pi}$ denotes the dissipative part of the pressure tensor, $\boldsymbol{\nabla}$ the gradient, ${\bf v}$ the velocity field, $\eta$ the shear viscosity, $\zeta$ the dilatational or bulk viscosity, $\boldsymbol{\mathcal J}_q$ the heat flux, $\kappa$ the heat conductivity, $\boldsymbol{\mathcal J}_k$ the diffusion flux of species $k$, ${\cal D}_k$ its diffusion coefficient, $\boldsymbol{\mathcal J}_{\rm e}$ the electric flux, $\boldsymbol{\sigma}$ the electric conductivity tensor, $\boldsymbol{\mathcal E}$ the electric field, $V$ the voltage, $R$ the resistance, and $I$ the electric current.

These processes are characterized by phenomenological laws (also called constitutive relations) and the associated parameters that are the transport coefficients and the reaction rate constants.

\vspace{-0.3cm}

\subsection{The contribution of a reaction to entropy production}

\vspace{-0.3cm}

Inspired by the works of Gibbs on thermodynamics and using the aforementioned concept of ``uncompensated heat'', De~Donder showed how chemical reactions contribute to entropy production \cite{D1936,R24}.  He introduced the concept of affinity for a reaction, defined as minus the change $\Delta G_r$ of Gibbs free energy in the reaction:
\be\label{affinity}
A_r \equiv - \Delta G_r = - \sum_k \mu_k \, \nu_{kr} \, ,
\ee
where $\mu_k$ is the chemical potential of species $k$ and $\nu_{kr}$ its stoichiometric coefficient in the reaction $r$.  Afterwards, in 1932, he established that the entropy production rate of a reaction is given by
\be
\frac{d_{\rm i}S}{dt}\Big\vert_r = \frac{1}{T} \, A_r \, W_r
\ee
in a well-mixed system, where $W_r=\int_V w_r \, d{\bf r}$ is the rate of the reaction in the volume $V$ of the system \cite{D1936}.  The important novelty is that the reactions (as well as the other rate processes) contribute by differences of chemical potentials, which could be uniform across the system, although the transport properties require gradients to drive the system away from equilibrium.

\vspace{-0.4cm}

\section{The non-equilibrium thermodynamics of reactive fluids}
\label{sec:noneq-thermo}

\vspace{-0.3cm}

The thermodynamics of continuous media was finally established around WWII on the following assumptions.  At the macroscale, we observe the slowest motions ruled by the fundamental conservation laws of energy, momentum, and mass; and also the conservation of particle numbers in the absence of any reaction, or the balance of particle numbers in the presence of slow enough reactions.  Furthermore, on long time scales, every element of matter is locally at equilibrium, so that the entropy density $s$ satisfies the following local Gibbs relation:
\be
ds=\frac{1}{T}\, de - \sum_k \frac{\mu_k}{T} \, dn_k \, ,
\ee
where $e$ is the density of internal energy and $n_k$ the particle density of species $k$.  Subsequently, in reactive fluids, the local balance equation for entropy~(\ref{local_2nd_law}) can be deduced from the local conservation and balance equations for the other quantities, showing that the entropy flux is given by
\be
\boldsymbol{\jmath}_s = s\, {\bf v} + \frac{1}{T} \, \boldsymbol{\mathcal J}_q - \sum_k  \frac{\mu_k}{T} \, \boldsymbol{\mathcal J}_k
\ee
in terms of the heat and diffusion fluxes, and the entropy production rate density by
\be
\sigma_s = \sum_\alpha {\mathcal A}_\alpha \, {\mathcal J}_\alpha \ge 0
\ee
in terms of the thermodynamic forces and fluxes associated with each irreversible process (see Table~\ref{Tab1}) \cite{E1940a,E1940b,E1940c,P47,P55,GM62,C65}.

\vspace{-0.4cm}

\begin{table}[h]
\caption{The thermodynamic forces ${\cal A}_a$ and fluxes ${\cal J}_a$ associated with the different irreversible processes of a reactive fluid. $T$~denotes the temperature, $\boldsymbol{\nabla}$ the gradient, ${\bf v}$ the velocity, $\boldsymbol{\Pi}$ the tensor of viscous pressure, $\boldsymbol{\mathsf 1}$ the unit tensor, $\boldsymbol{\mathcal J}_{q}$ the heat flux, $\mu_k$ the chemical potential of species $k$, $\boldsymbol{\mathcal J}_{k}$ its diffusion flux, $\nu_{kr}$ its stoichiometric coefficient in reaction $r$, and $w_r$ the rate density of the reaction.}
\label{Tab1}
\vskip 0.1 cm
\begin{tabular}{|lll|}
\hline
irreversible process & thermodynamic force ${\cal A}_{\alpha}$  & flux ${\cal J}_{\alpha}$\\
\hline \\[-10pt]
shear viscosity & $-\frac{1}{2T} \big(\boldsymbol{\nabla}{\bf v}+\boldsymbol{\nabla}{\bf v}^{\rm T}-\frac{2}{3}\boldsymbol{\nabla}\cdot{\bf v}\, {\boldsymbol{\mathsf 1}}\big)$ & $\boldsymbol{\Pi}-\Pi\, \boldsymbol{\mathsf 1}$ \\[2pt]
bulk viscosity &  $-\frac{1}{T}\, \boldsymbol{\nabla}\cdot {\bf v}$ & $\Pi = \frac{1}{3}\, {\rm tr}\,\boldsymbol{\Pi}$ \\[2pt]
heat conduction & $\boldsymbol{\nabla}(1/T)$ & $\boldsymbol{\mathcal J}_{q}$ \\[2pt]
diffusion & $\boldsymbol{\nabla}(-\mu_k/T)$ & $\boldsymbol{\mathcal J}_{k}$ \\[2pt]
reaction & $ - \frac{1}{T}\sum_{k} \mu_k \, \nu_{kr}$ & $w_{r}$ \\[2pt]
\hline
\end{tabular}
\end{table}

These results should be supplemented by non-equilibrium constitutive relations between the fluxes and the thermodynamic forces:
\be
{\cal J}_\alpha = \underbrace{\sum_\beta {\cal L}_{\alpha\beta}\, {\cal A}_\beta}_{\rm linear \, response} + \underbrace{\frac{1}{2} \sum_{\beta\gamma} {\cal M}_{\alpha\beta\gamma} \,{\cal A}_\beta\, {\cal A}_\gamma + \cdots}_{\rm nonlinear \, responses} \, ,
\ee
which must satisfy the Onsager reciprocal relations ${\cal L}_{\alpha\beta}={\cal L}_{\beta\alpha}$ because of microreversibility \cite{O31a,O31b} and the Curie symmetry principle holding in the continuous medium of interest \cite{C1894}.  These constitutive relations are linear if the mean free path is smaller than the length scales of variation in the free-energy landscape (which is the case for the viscosities, heat conduction, and diffusion); or nonlinear if the free-energy landscape varies over length scales of the order of molecular sizes (e.g., for chemical reactions).  In the latter case, the reaction rate density of a reaction is given by $w_r=w_{+r} \left( 1 - {\rm e}^{-A_r/k_{\rm B}T}\right)$ in terms of the affinity~(\ref{affinity}) according to the mass action law holding in dilute solutions.

\vspace{-0.4cm}

\section{General evolution criterion and dissipative structures}
\label{sec:DS-GEC}

\vspace{-0.3cm}

In 1954, Glansdorff and Prigogine obtained the general evolution criterion, which says that, ``in a system evolving in a locally stable thermodynamic phase and subjected to stationary constraints, the thermodynamic forces always change with time by decreasing the entropy production rate'' \cite{GP54}.

As aforementioned, the entropy production rate is defined by $P\equiv d_{\rm i}S/dt=\int_V \sigma_s\, d{\bf r}= \int_V \sum_\alpha {\mathcal A}_\alpha \, {\mathcal J}_\alpha \, d{\bf r}\ge 0$.  Its time derivative can thus be written as
\bea
&&\frac{dP}{dt} = \frac{d_{\boldsymbol{\mathcal A}}P}{dt} + \frac{d_{\boldsymbol{\mathcal J}}P}{dt}
\qquad\mbox{with}\qquad\nonumber\\
&&
\frac{d_{\boldsymbol{\mathcal A}}P}{dt} \equiv \int_V \sum_\alpha \frac{\partial{\cal A}_\alpha}{\partial t} \, {\cal J}_\alpha \, d{\bf r}
\qquad\mbox{and}\qquad\nonumber\\
&&
\frac{d_{\boldsymbol{\mathcal J}}P}{dt} \equiv \int_V \sum_\alpha {\cal A}_\alpha \, \frac{\partial{\cal J}_\alpha}{\partial t} \, d{\bf r} \, .
\eea
In systems without convective processes and subjected to the stated conditions, the following inequality is satisfied:
\be
\frac{d_{\boldsymbol{\mathcal A}}P}{dt} \le 0 \quad\left(\mbox{with}\ \ \frac{d_{\boldsymbol{\mathcal A}}P}{dt}=0 \ \ \mbox{in steady states}\right) ,
\ee
hence the general evolution criterion.  The criterion can be extended to include the convective processes \cite{GP54,GP71}.

If and only if, linear relations hold between the fluxes and the thermodynamic forces, the general evolution criterion implies the theorem of minimum entropy production, according to which
\be
\frac{d_{\boldsymbol{\mathcal A}}P}{dt} = \frac{d_{\boldsymbol{\mathcal J}}P}{dt}  =\frac{1}{2} \, \frac{dP}{dt}  \le 0 \, .
\ee
However, such linear relations do not hold in general, leading to the possibility of spatiotemporal pattern formation under far-from-equilibrium conditions.  The general evolution criterion can be used to determine the stability or instability of non-equilibrium steady states, showing that instabilities may happen far enough from equilibrium.  Beyond the threshold of instability, dissipative structures may thus emerge, in particular, in diffusion-reaction systems \cite{PN67,PL68,NP77,G18}.

\vspace{-0.4cm}

\section{The non-equilibrium thermodynamics of active matter}
\label{sec:AM}

\vspace{-0.3cm}

Significant advances have recently been made in research on the dissipative structures and the thermodynamics of active matter \cite{MJRLPRA13,K13,BDLRVV16,JGS18,GK19,GK20,MFTC21,ATR25,BRS25,G25}.  There exist many different sorts of active matter, which may be composed of self-propelled particles or microorganisms, locally converting energy into locomotion.  An example is given by active suspensions of micrometric Janus particles with a catalytic hemisphere, where some reactant is consumed generating propulsion by self-diffusiophoresis \cite{K13,GK19}.  In such systems, the internal degrees of freedom of the particles like those describing their orientation may also have slow motions.  In this respect, the local conservation equation of the Janus particles should be generalized to their distribution function $f_{\rm C}({\bf r},\boldsymbol{\alpha},t)$ defined in the six-dimensional space of their position~${\bf r}$ and orientation~$\boldsymbol{\alpha}$:
\be
\partial_t \, f_{\rm C} +\pmb{\nabla}\cdot(f_{\rm C} \, {\bf v} + \pmb{\mathscr J}_{\rm Ct}) + {\rm div}_{\rm r}(f_{\rm C} \, \pmb{\upsilon}_{\rm r} + \pmb{\mathscr J}_{\rm Cr}) = 0 \, ,
\ee
where ${\rm div}_{\rm r}$ denotes the divergence in the orientation subspace \cite{GK20,G25}.
In addition, the local Gibbs relation should be extended into
\be
\delta s = \frac{1}{T} \, \delta e - \sum_k \frac{\mu_k}{T} \, \delta n_k - \int \frac{\zeta_{\rm C}}{T} \, \delta f_{\rm C} \, do \, ,
\label{Gibbs-AM}
\ee
where $\zeta_{\rm C}$ is the chemical potential of the Janus particles and $do$ the element of integration in their orientation subspace \cite{PM53,P55,GM62}.  In such systems, the balance equation for entropy~(\ref{local_2nd_law}) can also be deduced and, consistently, the entropy production rate density should include contributions from the Janus particles:
\be
\sigma_s = \sum_\alpha {\mathcal A}_\alpha \, {\mathcal J}_\alpha + \int \sum_{\gamma={\rm t},{\rm r}} \pmb{\mathscr A}_{{\rm C}\gamma} \cdot \pmb{\mathscr J}_{{\rm C}\gamma} \, do \ge 0
\ee
with corresponding thermodynamic forces $\pmb{\mathscr A}_{{\rm C}\gamma}$ \cite{GK20,G25}.

\vspace{-0.4cm}

\section{The microscopic foundations of non-equilibrium thermodynamics}
\label{sec:MicroFoundations}

\vspace{-0.3cm}

Macroscopic theories like hydrodynamics and thermodynamics find their origins in the microscopic motion of atoms composing matter.  In particular, the Hamiltonian microdynamics of atoms implies the conservation of energy, which justifies the first law of thermodynamics.  However, entropy is a quantity characterizing the statistical disorder associated with a probability distribution defined in the space of microstates of the system, as shown by Boltzmann~\cite{B1877}.  Therefore, the second law of thermodynamics should be justified on the basis of statistical mechanics.

The earliest justification is based on the kinetic theory of gases and Boltzmann's kinetic equation for the time evolution of the distribution function $f({\bf r},{\bf p},t)$ of the positions ${\bf r}$ and momenta ${\bf p}$ of the particles in a dilute gas \cite{B1872}:
\be
\partial_t f + \frac{\bf p}{m}\cdot\boldsymbol{\nabla} f = \int \frac{d{\bf p'}}{m}\, d\Omega\, \sigma_{\rm diff}\, \Vert{\bf p}-{\bf p'}\Vert \, (\tilde f \tilde f' - f f') \, ,
\ee
where $m$ is the mass of the particles, $\sigma_{\rm diff}$ the differential cross-section of the binary collisions, $f$ \& $f'$ the distribution functions of the two particles of momenta ${\bf p}$ \& ${\bf p'}$ incoming a collision, $\tilde f$ \& $\tilde f'$ those of the two outgoing particles, and $d\Omega$ an element of solid angle for the collision direction.  If Boltzmann's equation holds, the second law of thermodynamics is satisfied according to the famous $H$-theorem.  Indeed, in dilute gases, the entropy is given by $S=-k_{\rm B}H=\int s \, d{\bf r}$, where the entropy density $s({\bf r},t)=k_{\rm B}\int d{\bf p} \, f({\bf r},{\bf p},t) \, \ln\frac{\rm e}{h^3f({\bf r},{\bf p},t)}$\footnote{$k_{\rm B}$ and $h$ are Boltzmann's and Planck's constants, respectively, and ${\rm e}=2.71828\dots$.} obeys the local balance equation~(\ref{local_2nd_law}) with an entropy production rate density $\sigma_s$, which can be proved to be always non-negative, fully justifying the second law in this framework.  

As a consequence of the $H$-theorem, in an isolated system, the relaxation towards equilibrium proceeds in two stages, as shown by Chapman and Enskog in 1917 \cite{CC39}.  Starting from an arbitrary initial distribution function $f({\bf r},{\bf p},0)$, first, there is a fast kinetic relaxation towards local equilibrium over a timescale of the order of a few intercollisional times and, afterwards, a slow hydrodynamic relaxation towards global equilibrium, which is asymptotically reached in the long-time limit.  In this second stage, the distribution function can be assumed to have a Maxwellian form at the local temperature, particle density, and velocity for every element in the fluid.  Therefore, the macroscopic hydrodynamic equations and the values of the transport coefficients can be deduced from the Boltzmann kinetic equation in dilute gases.  In this way, the microscopic foundations of non-equilibrium thermodynamics can be established for dilute gases and also for reactive gases \cite{P49} and plasmas \cite{B75}.

These results have been extended to condensed matter with the advent of the methods of non-equilibrium statistical mechanics for many-body systems. In the 1950s and early 1960s, Green~\cite{G52,G54}, Kubo~\cite{K57}, Mori~\cite{M58}, McLennan~\cite{McL63}, and others obtained formulas for the transport coefficients expressed in terms of microscopic global currents defined for the many-body system. These methods apply to the different phases of condensed matter \cite{MG20,MG23}.  More recently, the so-called fluctuation theorems have provided further justifications for the non-negativity of the entropy production \cite{G22}.  In addition, the typical trajectories of non-equilibrium stochastic processes have been shown to be more probable than their time reversal, meaning that directionality -- i.e., dynamical order -- manifests itself away from equilibrium, as a corollary of the second law of thermodynamics \cite{G22}.

Nowadays, we understand that microreversibility -- i.e., the symmetry of the Hamiltonian microdynamics under time reversal -- is always satisfied, but the time-reversal symmetry is broken in the statistical description of non-equilibrium systems.  Energy dissipation is explained by the fact that, since the fundamental interactions of nature are local, energy goes from few to many particles after several successive collisions and, thus, from the macroscale down to the microscopic degrees of freedom, generating heat, which is a degraded form of energy.  In this regard, the description of macroscopic systems requires coarse graining the microscopic variables, as carried out in non-equilibrium statistical mechanics \cite{G22}.

\vspace{-0.4cm}

\section{Conclusion and perspectives}
\label{sec:conclusion}

\vspace{-0.3cm}

Thermodynamics is the theory of heat and temperature.  It is based on the concept of entropy, which evaluates the amount of disorder in the microscopic degrees of freedom that are not under macroscopic control.  Thermodynamics distinguishes between reversible and irreversible processes and, thus, allows us to identify the irreversible processes in natural phenomena.
Non-equilibrium thermodynamics is known for reactive fluids, active matter, and other phases of matter like crystals, liquid crystals, or superfluids.

Dissipative structures such as oscillations or spatiotemporal patterns can emerge in far-from-equilibrium open systems, which is consistent with the thermodynamics of open systems.

Furthermore, the domains of validity of thermodynamics have been delimited by much effort in classical and quantum statistical mechanics.

Thermodynamics provides the framework to understand the mechanisms of energy conversion and their efficiency in engineering, the physical sciences, and the biosciences \cite{H77}.  This concerns, in particular, the biological systems, where the metabolism constitutes a chemical reaction network functioning away from equilibrium.  Moreover, as envisaged by Brillouin~\cite{B56} and Landauer~\cite{L61}, thermodynamics plays a fundamental role in the relationship between energy and information, which concerns information science, microelectronics, and also molecular biology in the non-equilibrium processes of replication, transcription, and translation, where connections between thermodynamics and information theory have been established \cite{AG08}.

\vspace{-0.4cm}
\section*{Acknowledgments}
\vspace{-0.3cm}

This research was supported by the Universit\'e Libre de Bruxelles (ULB).  This presentation was given at the Belgian Symposium of Thermodynamics ``Carnot 2024'', 16-18 December 2024, Li\`ege, Belgium.


\end{document}